\documentstyle[aasms4,psfig]{article}
\begin{document}

\def\simlt{\lower.5ex\hbox{$\; \buildrel < \over \sim \;$}}
\def\simgt{\lower.5ex\hbox{$\; \buildrel > \over \sim \;$}}
\title{X-rays and Protostars in the Trifid Nebula}
\author{Jeonghee Rho}
\affil{Infrared Processing and Analysis Center, California Institute of
Technology, MS 100-22  Pasadena, CA 91125, rho@ipac.caltech.edu}
\author{Michael F. Corcoran}
\affil{Code 662, NASA/Goddard Space Flight Center,
Greenbelt, MD 20742}
\author{You-Hua Chu}
\affil{Department of Astronomy, University of Illinois, 1002 West Green
Street, Urbana, IL 61801}
\centerline{AND}
\author{William T. Reach}
\affil{SIRTF Center, California Institute of
Technology, MS 100-22  Pasadena, CA 91125}

\vskip 2truecm

\received{February 19 2001}
%\accepted{}

\begin{abstract}

The Trifid Nebula is a young HII region recently rediscovered as a
``pre-Orion" star forming region, containing protostars undergoing
violent mass ejections visible in optical jets as seen in images from
the Infrared Space Observatory and the Hubble Space Telescope.  We
report the first X-ray observations of the Trifid nebula using ROSAT
and ASCA.  The ROSAT image shows a dozen X-ray sources, with the
brightest X-ray source being the O7 star, HD 164492, which provides
most of the ionization in the nebula. We also identify 85 T Tauri star
and young, massive star candidates from near-infrared colors using the
JHK$_s$ color-color diagram from the Two Micron All Sky Survey
(2MASS).  Ten X-ray sources have counterpart near-infrared sources. The
2MASS stars and X-ray sources suggest there are potentially numerous
protostars in the young HII region of the Trifid.   ASCA moderate
resolution spectroscopy of the brightest source shows hard emission up
to 10 keV with a clearly detected Fe K line. The best model fit is a
two-temperature ($T = 1.2\times 10^{6}$ K and $39\times 10^{6}$ K)
thermal model with additional warm absorbing media.  The hotter
component has an unusually high temperature for either an O star or an
HII region; a typical Galactic HII region could not be the primary
source for such hot temperature plasma and the Fe XXV line emission.
We suggest that the hotter component originates in either the interaction
of the wind with another object (a companion star or a dense region of
the nebula) or from flares from deeply embedded young stars.

%Trifid Nebula.
\noindent{\it Subject headings}: HII regions - Stars: formation -
X-rays: individual (Trifid Nebula) - infrared:Stars
\end{abstract}

\section{Introduction}

\ion{H}{2} regions contain various types of X-ray emitting sources such as
high- and low-mass stars, binaries, and protostars.  Single massive O
and B stars emit soft X-rays from shocks in their radiatively
unstable, outwardly moving outer atmospheres.  X-ray emission from
late-type stars is attributed to magnetically heated stellar coronae
and magnetically-driven stellar flares.  Protostars produce X-rays via
magnetic heating, and perhaps from accretion or excretion disks and/or
interaction of stellar jets with the circum-nebular gas.  Massive
binaries can produce X-ray emission from colliding stellar winds;
close low-mass binaries can produce X-rays via mass exchange and
enhanced dynamo action.  X-ray emission provides a sensitive tracer of
these different stellar populations, and along with infrared colors,
is one of the main probes to identify protostars and pre-main-sequence
(PMS) stars.

Studies of X-ray emission from \ion{H}{2} regions/star forming regions have
recently made significant advances due to ROSAT (Casanova et al.\ 1995), 
ASCA (e.g., Koyama et al.\  1996), and most recently
{\it Chandra} observations (Garmire et al.  2000).  For example, in
the $\rho$ Oph core region, 70\% of the near-infrared sources associated
with protostars and molecular cores have X-ray counterparts (Casanova
et al.\ 1995), and a large number of X-ray sources are found to be
low-mass PMS stars in other star forming
regions such as Orion (\cite{Alcala97}), Chamaeleon
(\cite{Alcala96}; Feigelson et al.\ 1993), Lupus (Krautter et al.\ 1997),
and Taurus-Auriga (Neuh\"user et al.\ 1995; Wichmann et al.\ 1997).
Flare-like X-ray events have been detected from T Tauri stars (Kamata
et al.\ 1997; Koyama et al.\ 1994).  The X-ray emitting sources in the
Monoceros and Rosette molecular clouds are also mostly T Tauri and
Herbig Ae/Be stars and they typically show luminosities of L$_{\rm
x}$$\sim$10$^{30}$--10$^{32}$ ergs s$^{-1}$ (Gregorio-Hetem et al.\ 1998).
Recent {\it Chandra}
observation of the Orion Nebula resolved a thousand X-ray emitting
PMS stars with a mass range of 0.05 M$_{\odot}$ to 50
M$_{\odot}$, and a combined infrared and X-ray study suggested that the
X-ray luminosity depends on stellar mass, rotational history, and
magnetic field (Garmire et al.\ 2000).

At a distance of 1.67 kpc (Lynds et al.\ 1985), the Trifid Nebula,
M~20, is one of the best-known astrophysical objects and one of the
prettiest: it glows brightly in red light and is trisected by
obscuring dust lanes.  At an age of $\sim 3\times 10^{5}$ years, the
Trifid is one of the youngest known \ion{H}{2} regions.  Observations
with the Infrared Space Observatory (ISO) and the Hubble Space
Telescope (HST) show the Trifid as a dynamic, ``pre-Orion" star
forming region containing young stars undergoing episodes of violent
mass ejections, and protostars (like HH399) losing mass and energy to
the nebula in optically bright jets (Cernicharo et al.\ 1998,
hereafter CLC98; Lefloch \& Cernicharo 2000; Hester et al.\ 1999).
The ionization of the nebular gas is dominated by the O7.5 star HD
164492.  HD 164492 is a luminosity class V (Levato 1975) or III (Conti
\& Alschuler 1971) star with a bolometric luminosity of L$_{bol}$
$\sim$0.5--1.6$\times10^{39}$ ergs s$^{-1}$ and an X-ray luminosity of
6$\times10^{32}$ ergs s$^{-1}$ (Chlebowski et al.  1989).  The mass
loss rate of this O star is $\dot M$ = 2$\times10^{-6}$ M$_{\odot}$
yr$^{-1}$ (Howarth \& Prinja 1989) and the wind terminal velocity is
$V_{\infty}$ = 1580 km s$^{-1}$ (Prinja, Barlow, \& Howarth 1990).

X-ray observations of the Trifid offer a unique opportunity to study
the influence of a massive star on star formation in an exceptionally
young star forming region.  Initially X-ray emission from the Trifid
was reported only from the O star in the {\it Einstein} IPC catalog
(Chlebowski et al.\ 1989).  We serendipitously discovered a complex of
X-ray emission from the Trifid Nebula in a PSPC observation of the
nearby supernova remnant W28 (Rho et al.\ 1995).  Subsequently, we
started an extensive investigation of the X-ray emission from the
Trifid Nebula.  In this paper, we present the first detection of a
dozen X-ray sources in the Trifid Nebula, and we correlate these with
protostar candidates identified using the Two Micron All Sky Survey
(2MASS) data.  The ROSAT images show multiple point sources including
HD 164492 and several T Tauri stars, and the ASCA spectra
show hard X-ray emission including detection of an Fe K line.  We
discuss identifications of the X-ray emitting sources and the origin
of the unusually hard X-ray emission from the Trifid.

\section{X-ray Sources in the Trifid}

\subsection{X-ray Observations}

The Trifid Nebula was observed using the X-ray telescope on ROSAT
(Tr\"{u}mper 1993) with the Position Sensitive Proportional Counter
(PSPC) as the imaging detector.  The PSPC on-axis angular resolution
is 25$^{\prime\prime}$ (FWHM, at 1 keV), and the PSPC covers a
2$^{\circ}$ field of view in the 0.1--2.4 keV energy band.  Two PSPC
observations were analyzed for this paper: rp900375 centered on HD
164492 and observed on 1993 September 8 for an exposure of 9,365 s
(PI: S.\ Snowden), and rp500236 centered on the supernova remnant W28
and observed on 1993 April 1 for an exposure of 10,476 s (PI: R.\
Pisarski; Rho et al.\ 1995).

We also performed an ASCA observation (PI: J.\ Rho; sequence number
26051000) toward the center of the Trifid Nebula.  The observation
took place on 1998 September 30 to October 2.  ASCA (Tanaka 1992) has
two detector pairs: Gas Imaging Spectrometers (GIS2 and GIS3) and
Solid-state Imaging Spectrometers (SIS0 and SIS1).  The SIS covers an
energy band of 0.5--10 keV and the GIS 0.6--10 keV. The on-axis
angular resolution of the GIS and SIS is about 1--2 arcminutes.  Each
GIS counter has a circular field of view of 35$'$ diameter while the
field of view of each SIS CCD is an 11$'$ square, and thus both GIS
and SIS detectors sufficiently cover the entire Trifid Nebula.  We
filtered the data using a few criteria such as Cut off Rigidity (COR)
and earth elevation (based on Revison 2 processing).  After filtering,
the exposure time was 57 ks for the GIS and 53.5 ks for the SIS. The
entire Trifid Nebula region after background subtraction has count
rates of 0.050$\pm$0.001 cts s$^{-1}$ for SIS0, 0.040$\pm$0.001 cts
s$^{-1}$ for SIS1, 0.030$\pm$0.008 cts s$^{-1}$ for GIS2,
0.039$\pm$0.009 cts s$^{-1}$ for GIS3, and 0.086$\pm$0.004 cts
s$^{-1}$ for the ROSAT PSPC, in their respective energy bands
(integrated over all channels).

\subsection{X-ray Source Identification}

The ROSAT PSPC image of the Trifid Nebula is shown in Figure 1.  This
image for the first time reveals that the Trifid Nebula contains
numerous X-ray sources.  We have identified X-ray sources in the PSPC
image using the FTOOLS task SRCDETECT and estimated the count rates
and uncertainties.  The detected point sources are presented, in order
of increasing right ascension, in Table ~\ref{trixray} and marked in
Figure 1.  Table ~\ref{trixray} lists the position, count rate, and
$\sigma$ of detection for ten sources detected at $>3 \sigma$ and two
possible sources (sources 11 and 12) detected with somewhat lower
confidence.  We here define new X-ray sources detected in ROSAT PSPC
image, as ROSAT X-ray source in the Trifid (RXT).  Since the count
rate is very small except for the O star HD 164492, we estimated the
luminosity by assuming an absorption column density N$_H$ =
3$\times$10$^{21}$cm$^{-2}$ (A$_{\rm V}$$\sim$ 1.5$^{m}$, see Section
4 for details), and a thermal spectrum with $kT$=1 keV. The X-ray
emission of PMS stars is understood to be thermal emission from gas
rapidly heated to a temperature of $\sim$1 keV by violent
magnetohydrodynamical reconnection events (Feigelson \& Montmerle
1999).  The correspondence between the PSPC count rate and the X-ray
unabsorbed flux is 1$\times10^{-3}$ PSPC cts s$^{-1}$ $\sim$
2.95$\times$10$^{-14}$ ergs s$^{-1}$ cm$^{-2}$.  Using this
conversion, the luminosities of the X-ray sources are computed and
given in Table ~\ref{trixray}.

We have examined the source list database using SIMBAD, identified
counterparts to the X-ray sources at other wavelengths, and marked
them in Figure 1.  To visualize the correspondence of X-ray sources
with either optical or radio sources, we have plotted the X-ray
contours over an H$\alpha$ image (F. Winkler, private
communication) in Figure 1.  Twenty-four sources from the Guide Star
Catalog (GSC) are visible in the H$\alpha$ image in Figure 2.  The
brightest X-ray point source, RXT8 in Table ~\ref{trixray},
corresponds to the O star HD 164492.  The possible X-ray source RXT11
coincides with the B8 star HD 313596 (R.A.\ $18^{\rm h} 02^{\rm m}
35^{\rm s}$ and Dec.\ $-22^\circ$59$^{\prime} 54^{\prime \prime}$),
and the possible X-ray source RXT12 with the optical star
GSC\_06842\_00001 (R.A.\ $18^{\rm h} 02^{\rm m} 34.8^{\rm s}$ and
Dec.\ $-23^\circ$03$^{\prime} 06.3^{\prime \prime}$).  None of the
other GSC stars coincide with X-ray peaks.  A radio source, GPSR5
6.980$-$0.286 (R.A.\ $18^{\rm h} 02^{\rm m} 28.1^{\rm s}$ and Dec.\
$-23^\circ$03$^{\prime} 46.3^{\prime \prime}$; Becker et al.\ 1994),
is close to the X-ray emitting area, but does not have a corresponding
X-ray peak.  Four protostars (TC0, TC1, TC3 and TC4 sources
\footnote {The source designation TC0, TC1, etc. was assigned by CLC98. Since this
designation was already in use, for indexing purposes these sources 
should be referred to as [CLC98] 0, etc.}
in CLC98)
 have been reported in the Trifid Nebula (CLC98),
which are marked in
Figure 1, but only one is close to the X-ray peak at HD 164492.  In
the next section we correlate the PSPC X-ray sources with sources
showing near-infrared color excesses, and present candidate
protostars.

\section{Near-Infrared Sources from 2MASS: Young Stellar Objects }

We have identified Young Stellar Objects (YSOs) using the Two Micron All Sky
Survey (2MASS) data (\cite{Skrutskie}). Using identical telescopes in the
northern and southern hemispheres, 2MASS is mapping the entire sky in
the J (1.11-1.36$\mu$m), H (1.5-1.8$\mu$m) and K$_s$ (2-2.32$\mu$m)
bands to a limiting point source sensitivity of approximately 16.5 mag,
16.0 mag, and 15.5 mag, respectively (\cite{Cutri}).  The data
towards the Trifid Nebula were taken on 1998 June 14 using the
southern telescope, and most of these data were included in the 2MASS Second
Incremental Release, but  a small portion of area was
in the 2MASS Working Database due to large photometric uncertainties 
at the time of
the Incremental Release.
The photometry is typically better than 5\%
(Cutri et al. 2000).

We used the 2MASS point source catalog to extract
sources within an 8-arcmin radius centered on R.A.\ $18^{\rm h} 02^{\rm m}
30^{\rm s}$ and Dec.\ $-23^\circ$02$^{\prime} 00^{\prime \prime}$.
We have selected sources with the following
criteria. First, we selected the sources which were detected in all
three J, H, and K$_s$ bands,
We then selected sources with the signal to noise ratio greater than
10 (i.e., the J, H and K$_s$ magnitudes are brighter than 15.8 mag,
15.1 mag, and 14.3 mag, respectively).  These selections produced
$\sim$1,100 such sources.  We then accepted sources with photometric
uncertainties $\sigma< 0.25$ mag whose fit to the point spread
function produced reduced $\chi^2_{\nu}<2$.  This last criterion
excluded blended sources that caused higher uncertainties in the
photometry.  This is important in the Galactic plane where
near-infrared sources are crowded and the 2MASS has a limited spatial
resolution (3\farcs5).  This criterion removed inaccurate blue points
that appeared in the JHK$_s$ color-color diagram (as described below).
The criteria we have used are conservative for the magnitude limit and
photometric uncertainties of the 2MASS survey.

We plotted the sources in the JHK$_s$ color-color diagram, as shown in
Figure 3a, in order to identify YSOs with infrared color excess. Figure
3b shows H-K$_s$ vs K$_s$ magnitude for the selected sample.  The
interstellar reddening vector from Rieke \& Lebofsky (1985) is also
plotted.  Adopting the intrinsic colors of giant and dwarf stars from
Bessell and Brett (1988), we find the visual extinction towards this
direction is as high as A$_{\rm V}$=30 mag for stars on the back side
of the Trifid Nebula. The observed JHK$_s$ colors of YSOs can be
explained by circumstellar disk models (Lada \& Adams 1992).  We
identified the T Tauri stars with (J-H)$_{CTTS}$ =
0.58$\pm$0.11$\times$(H-K$_{s}$)$_{CTTS}$ +0.52$\pm$0.06 (Meyer et al.
1997) and H-K$_s$ $>$0.6 mag, where CTTS is classical T Tauri stars.
These T Tauri stars fall between the two solid lines in Figure 3a. The
stars below the extinction curve (dashed line in Figure 3a) and above
the T Tauri star lines (solid line in Figure 3a) are massive YSOs (Lada
\& Adams 1992); these stars are plotted as diamonds in Figure 3a.  In
total, we found 41 T Tauri star candidates and 44 massive YSO
candidates from the 2MASS sources which are listed in Table 
~\ref{tri2mass}. The T Tauri stars and YSOs are generally located along
the ionization front. Figure 3c shows optical images, and T Tauri
stars, massive YSOs, and X-ray sources are marked. Figure 3d shows a
color composite of the J, H and K$_s$ images, with the J image in blue,
H image in green, and K$_s$ image in red. T Tauri stars, massive YSOs,
and X-ray sources are marked.  The image also shows a number of
protostars at the ionization front along the dust lane.  

These protostar candidates were cross-correlated with the X-ray
sources.  Taking into account the resolution of the PSPC, we identify
a possible coincidence if the separation between a 2MASS source and an
X-ray source is less than 15$''$.  Three X-ray sources (RXT 1, 6, and
7) are coincident with either T Tauri stars or YSOs; they are noted in
Table ~\ref{triinfmatch}.  Figure 3d shows that many red 2MASS stars
within the X-ray source error boxes could be considered coincidences,
but they are not included in the list of sources selected according to
the aforementioned conservative criteria, as they are not detected in
one or both of the J and H bands.  Therefore, we have relaxed the
selection criteria and extracted 2MASS sources from the same area
excluding only sources with an artifact flag.  We made a list of the
2MASS sources that are within 15$''$ from X-ray sources.  Each X-ray
source has $\sim$10 2MASS counterpart candidates; only the red stars
are identified as possible counterparts to the X-ray sources.  Seven
red 2MASS star counterparts of X-ray sources are identified and listed
in Table ~\ref{triinfmatch}, with J, H, and K$_s$ magnitudes and
reasons that they were not selected in the earlier list of T Tauri
star or massive YSO candidates (either J=`fil' or high reduced
$\chi^2$ indicating multiple or extended sources with large
photometric uncertainties).  These 7 sources are likely protostars
because their infrared colors are red and they also emit X-rays.  By
itself, an infrared color excess indicates either the presence of a
YSO or a heavily extincted main-sequence star.  The 2MASS counterparts
of the X-ray sources are marked in the JHK$_s$ color-color diagram
(Figure 3a) and on the 2MASS composite image (Figure 3d).  A few X-ray
sources do not have obvious optical or 2MASS counterparts.

There are five known massive YSOs in the Trifid with mid-infrared
emission detected using ISO (CLC98).
%these sources are marked
%with triangles in Figure 3b. 
These protostars were not identified as proto-stellar candidates from
the 2MASS data, probably because they are too deeply embedded to be
detected in the near-infrared.  A 2MASS red star is the counterpart of
the protostar TC2 in CLC98; we expect there are large numbers of such
embedded protostars in the Trifid Nebula not detected in
the near-infrared observations.  This population is likely correlated
with highly extincted dust lanes (see Figure 3d) and molecular clouds.

\section{X-ray Spectral analysis}

The ASCA SIS image is shown in Fig.  4 with the PSPC contours
superposed.  The X-ray emission is dominated by the emission from the
O star.  We extracted ASCA and ROSAT spectra from the entire region of
the Trifid Nebula.  The spectrum (Fig.  5a) shows clear Fe K line
emission along with weak Si and S lines, and a hard continuum tail up
to 10 keV, the highest energy observable by ASCA. We extracted a
background-corrected spectrum from a smaller region centered on HD
164492.  HD 164492 is the brightest X-ray source in the Trifid, and
the shape of the spectrum of the entire region in the ASCA data is not
significantly different from the spectrum of this star alone due to
the broad ASCA point spread function.  We made hard ($>$ 3keV) and
soft ($<$ 3keV) maps using the ASCA data, but no obvious difference
was noticeable at the spatial resolution (1$'$) of ASCA. We also
extracted an off-source ASCA spectrum in this direction, suspecting a
contribution from the Galactic ridge emission.  However, the
off-source spectrum showed that the observed off-source emission is
dominated by scattered emission from sources within the Trifid Nebula.
The detection of hard emission from the Trifid is unusual, because
most single O stars have very little emission at energies above 2 keV
and rarely show Fe K emission (\cite{Corcoran93}).  The only massive
stars to show such hard X-ray spectra are binaries, either high mass
X-ray binaries (HMXBs) with collapsed companions, or colliding wind
binaries with non-collapsed (O or WR star) companions which have
strong stellar winds and significant colliding wind X-ray emission.

We simultaneously fit the set of five spectra-- the ROSAT/PSPC,
ASCA/SIS0, SIS1, GIS2 and GIS3 spectra-- using single- or
two-temperature thermal models (Mewe-Kaastra plasma model; Kaastra
1992) with a single absorbing column density N$_H$.  The fits were
unacceptable (reduced $\chi^2$ of 3).  We next attempted a
two-temperature model.  Following Corcoran et al.'s (1994) models of
$\delta$ Ori and $\lambda$ Ori, we also included an additional ionized
(``warm") absorbing medium, as representative of the photoionized
stellar wind material (Waldron 1984; \cite{Corcoran94}), and allowed
different amounts of absorption for the hot and cold components.  The
line-of-sight extinction value is known toward this direction; A$_{\rm
V}$ = 1.3-1.5 mag, i.e., E(B-V) $\sim$ 0.3-0.4 mag (Kohoutek et al.\
1999; Lynds \& O'Neil 1985).  Using E(B-V) of 0.4 mag, we expect an
ISM N$_H$ of $\sim$3$\times$10$^{21}$ cm$^{-2}$, with which we have
fixed the N$_H$ value in our fit (also note that when we allow N$_H$
to vary, 3$\times$10$^{21}$ cm$^{-2}$ falls within the errors).  The
model yielded an acceptable fit with $kT_{1}$$\sim$0.14 keV
(1.2$\times10^6$ K) and N$_{H,1}=5.9\times$ 10$^{21}$ cm$^{-2}$,
and a hotter component with $kT_{2}$$\sim$3.3 keV (3.9$\times10^7$ K)
and N$_{H,2}= 2.7\times$ 10$^{21}$ cm$^{-2}$, with a line of sight
ISM N$_H$= 3$\times$10$^{21}$ cm$^{-2}$.  The abundances are fixed at
solar abundances.  The fit results are summarized in Table
~\ref{trispec}, along with the Fe K line characteristics.  The cold
component arises from the O star atmosphere but the hot component
might arise from a number of different sources such as unresolved
interacting binaries, active low mass stars, or PMS stars.

\section{The Nature of The Hard X-ray Component}

Our simultaneous ASCA/PSPC fits yield a total X-ray luminosity of
1.9-2.5$\times$$10^{34}$\,erg\,s$^{-1}$ (0.3-10 keV) using the
two-temperature model shown in Table ~\ref{trispec}.  If attributed to
HD 164492, then the ratio of X-ray and bolometric luminosities (=log
L$_{\rm x}$/L$_{bol}$) is between $-$5.0 and $-$4.5, which is much
higher than the typical ratios of log L$_x$/L$_{bol}$ $\sim$ $-$7 for
a single O-type star (Chlebowski 1989; Bergh\"ofer et al.\ 1996).  The
3 keV X-ray component is somewhat of a mystery, since winds from
single O-type stars are not known to produce such high temperature
emission.  Typically the highest temperature emission observed in O
star X-ray spectra has $kT < 1$ keV (e.g., Corcoran et al.  1994).

We discuss a few possibilities to explain the hot component in the
Trifid Nebula.  It may be that the hot component arises in the
interaction of the wind from the O7.5 star with another object (either
a companion star or a dense region of the nebula) outside the O star
atmosphere.  Colliding winds between an early-type star and an
early-type companion (another O star or a Wolf-Rayet star) can produce
shock-heated material in the wind interaction regions (Stevens et al.
1992) reaching temperatures of 10$^7-10^8$ K, and emit X-rays. 
% delete- redundant: In a
% colliding stellar wind model, X-ray emission is caused by gas heated
% to temperatures of 10$^7$-10$^8$ K behind the shock waves. 
However, recent photometric and spectroscopic studies of this region
found no evidence of a companion for HD 164492 (Kohoutek, Mayer \&
Lorenz 1999).  Unless the collision occurs far from the star ($d >
25''$) it will be unresolved to the ROSAT PSPC. However, recent radio
and near-infrared observations toward the central region of the Trifid
detected 3 sources close to the O star (Yusef-Zadeh et al.  2000)
which may be either stars or nebular knots photoionized by the UV
field of HD 164492.

Hard emission could also arise in single O stars from non-thermal
emission produced by Fermi acceleration by shocks in the O star wind
(Chen \& White 1991).  The hard component is suggested as a
non-thermal tail produced by inverse Comptonization of the
photospheric UV field by a population of fast particles accelerated by
a distribution of shocks.  Although it is not possible to determine if
the hard emission is from a non-thermal tail or the two thermal
temperature component from goodness of the fit to the spectra, the
presence of the Fe K line suggests that the second component is
thermal.  The sources of the two spectral components need to be
resolved spatially in order to determine their origins.  Enhanced hard
X-ray emission might also be produced by an oblique magnetic rotator
as suspected in another O7 star, $\theta^1$ Orionis C, the central star of the
Orion nebula (Gagn\'e et al.  1997).

The X-ray properties of the core of the HII region W3 share a number
of similarities to the Trifid emission: for W3, the luminosity is a
few 10$^{33}$ erg s$^{-1}$ with a similarly high temperature (Hofner
\& Churchwell 1997).  In W3 (and possibly in the Trifid), the high
temperature component may be produced by a hot, wind-shocked cavity
that results when strong stellar winds interact with a surrounding
dense molecular cloud (e.g. Churchwell 1990).  The presence of a known
young stellar object (TC1 in CLC98) is consistent with the presence of
a molecular cloud in the Trifid.  In addition, IRAS observations of HD
164492 (van Buren et al.  1995) show a bow-shock structure around the
star, perhaps indicative of a wind-cloud collision which might produce
the high temperature X-ray emission.  The stellar wind outflow at a
speed of 1600 km s$^{-1}$ should produce a post-shock temperature of
$\sim$30 million degrees, though the observed temperature may be
lower since radiative cooling is rapid.
CLC98 suggested that the HCO$^{+}$ molecular clouds
(likely the dust lanes) are fragmented shell around the nebula. In other
words, the clouds and dust lanes we see in the optical image are
located at the surface of the ionized sphere. If this is the case,
the 1600 km s$^{-1}$ wind will be interacting with the
lower-density, ionized medium and the shocked stellar wind
can emit at high temperatures, while the
photoionized materials of the edge of
clouds could supply sufficient density to emit strong X-rays.

The other possibility is that the hot component arises from deeply
embedded young stars especially since at least one embedded T Tauri
star (TC1 in CLC98, which is marked in Figure 1) exists near the O
star.  The TC1 source in CLC98 shows a large shift in the spectral
energy distribution and violent ejections of high-velocity material
(CLC98).  Other ASCA observations showed bright X-ray sources with
temperatures of 2-5 keV due to flares of protostars in the $\rho$
Ophiuchi dark cloud (Koyama et al.  1994), the Orion Nebula (Yamauchi
\& Koyama 1993), and the R Coronae Australis molecular cloud (Koyama
et al.  1996).  The hard emission was attributed to flares from
individual PMS stars with typical X-ray luminosities in the range
10$^{30-32}$ erg s$^{-1}$, and the peak luminosity of flares is shown
to be as large as 10$^{33-35}$ erg s$^{-1}$ and a temperature as high
as 10$^{8}$ K (\cite{Feigelsonm}; Grosso et al.  1997).  The hard
emission from W3 may be of a similar origin.  Signs of active star
formation in the Trifid have recently been reported (Lefloch et al.
2001): there is a dust cocoon or circumstellar disk around several
members in the center of the Trifid, and one young stellar source
shows a silicate feature in the circumstellar disk.  
Neither the $\rho$ Ophiuchi dark cloud or the R
Coronae Australis molecular cloud is as bright
in hard X-rays as the Trifid Nebula,  which may imply there are
higher number of protostars present in the Trifid. Large numbers of 
protostars unresolved to ASCA would dilute any flux variability 
produced by flares.

In summary, one of two scenarios is likely responsible for the hard
emission: the emission may arise from HD 169942 by the interaction of
the wind from the O star with another object (a companion star or a
dense region of the nebula), or from unresolved emission from active
PMS stars.  With our current data, we can not determine if one is more
favored.  To conclusively identify the hot component, a high
resolution image is needed to locate the emitting object in order to
determine whether the observed emission is produced near the O star,
or whether a distributed group of active PMS stars dominates the
observed emission.  Along with the images, time resolved spectra could
allow us to distinguish whether the hard emission is flare-like (time
variable).

\section{Identification of X-ray and Infrared Sources within the Trifid}

The detected X-ray sources and their counterparts are listed in Table
~\ref{triinfmatch}.  Most of X-ray sources are likely protostars or
PMS stars; one source is a T Tauri star, two are massive protostars,
and the others are unclassified protostars.  The JHK$_s$ color-color
diagram using the 2MASS data suggests that there are $\sim$80
protostars present in this nebula.  It has been already shown (CLC98;
\cite{Leoch}) that massive protostars (17-60 M$_\odot$) are forming in
the Trifid and they are associated with molecular gas condensations at
the edges of clouds, and their dynamical ages are 3$\times 10^4$ yr.
Whether low-mass protostars and T Tauri stars can be formed in a young
(3$\times$10$^5$ yr) region such as the Trifid is still an open
question.  Low-mass PMS stars of similarly young age were found in the
Orion Nebula using new {\it Chandra} observations (\cite{Hillenbrand};
\cite{Garmire}).  The {\it Chandra} observation also showed the
presence of young, low mass (0.1-3M$_\odot$) PMS stars as X-ray
sources (Garmire et al.  2000).  The populations of low mass and
massive protostars are similar in the Trifid, while we expect higher
populations of low mass protostars based on the initial mass function.
This is consistent with the fact that the Trifid is a very young HII
region; T Tauri stars have yet to form there.  The distribution of T
Tauri stars and massive YSOs is not obviously correlated with the
molecular cloud distribution.  It is possible that they are highly
embedded in the molecular clouds, and their near-infrared colors
cannot be fully obtained to identify protostars because either J
and/or H flux is unavailable.  This is consistent with the fact that
2MASS images show a higher population of red stars in the southern
part of the Trifid.  The HST images covering the southern part
suggested presence of embedded stars at the
head of the evaporating globules (Hester et al.  1999).  Deep
near-infrared images and spectroscopy will likely reveal hundreds of
young protostars in the Trifid as suggested by 2MASS and HST
data.

Whether diffuse X-ray emission exists within the Trifid Nebula is
currently unknown, because of the limited spatial resolution of the
PSPC images.  For the unidentified X-ray sources 5 and 10, we cannot
determine whether they are a part of diffuse emission or whether they
are real point sources.  They are very likely normal stars, but the
possibility that they are knots of diffuse emission can not be ruled out.
Diffuse X-ray emission from HII regions has been detected, although it
is rare.  A few examples are found such as in the Carina Nebula
(Seward \& Chlebowski 1982),
RCW 49 (Goldwurm et al.\ 1987; Belloni \& Mereghetti 1994), and the
Cygnus Superbubble (Bochkarev \& Sitnik 1985), and recently Wang
(1999) reported diffuse X-ray emission from the giant HII region 30
Dor in the Large Magellanic Cloud.  ROSAT and BBXRT observations of
the Carina Nebula show large-scale diffuse emission over at least
40$^{\prime}$, as well as discrete X-ray sources and hot gas
surrounding $\eta$ Car (\cite{Corcoran}).  The nature of diffuse
emission is unclear in HII regions.  Seward \& Chlebowski (1982)
suggested that stellar winds from the OB association adequately heat
the plasma. Wang (1999) suggested that the X-ray thermal
diffuse emission arises in blister-shaped region by loops of ionized gas
and the structure
is explained by the mass loading of the hot gas,
produced  by the central OB association.

We compare the Trifid Nebula with 30 Dor to determine whether stellar
winds in the Trifid may produce observable diffuse X-rays.  The stellar
wind luminosity in 30 Dor is a few $\times10^{39}$ ergs~s$^{-1}$, and
the X-ray luminosity of 30 Dor is $\sim$10$^{38}$ ergs~s$^{-1}$ (Wang
1999).  The stellar wind luminosity of HD~146692 is
$ {L_w} = 1.7\times 10^{36} ({\dot
M/{2\times10^{-6} {\rm M}_{\odot} yr^{-1}}} ) ({V/{1580\,{\rm
km}\,{\rm s}^{-1}}})^2 $ erg s$^{-1} $.
If the Trifid emits diffuse X-rays similarly to 30 Dor, we would expect
$\sim10^{35}$ ergs~s$^{-1}$ diffuse X-ray emission from the Trifid.  This
is higher than the total X-ray luminosity of the Trifid.  It is likely
that supernova heating contributes significantly to the bright X-ray
emission from 30 Dor.  The Trifid Nebula is too young to have hosted
supernova explosions, so its diffuse X-ray emission should be much
fainter than that of 30 Dor.

Future high resolution X-ray observations by new telescopes such as
\textit{Chandra} and \textit{XMM} should be able to resolve the O star
from its immediate environment and discrete X-ray sources such as T
Tauri stars, numerous protostars, young and old normal stars, and to
resolve the source of the high temperature emission.  A deep
near-infrared image with other wavelength observations can identify
the PMS stars and protostars, and their mass populations.  The Trifid
Nebula is an exciting laboratory to understand early stage of star
forming activities in HII region.

\acknowledgements 

We thank Lynne Hillenbrand for  helpful discussion on near-infrared
colors of protostars and protostar disk models, and for useful comments
on the manuscript, and John Carpenter for helpful discussion on 2MASS
data.  We thank Dr.  Frank Winkler for allowing us to reproduce his
optical image.  This work for J. R. is partially supported by NASA/ADP
grant, NASA-1407. This publication makes use of data products from the
Two Micron All Sky Survey, which is a joint project of the University
of Massachusetts and the Infrared Processing and Analysis Center,
funded by the National Aeronautics and Space Administration and the
National Science Foundation. J. R. and W. T. R. acknowledge the support
of the Jet Propulsion Laboratory, California Institute of Technology,
which is operated under contract with NASA.

\begin{table}
\caption{X-ray sources detected in the PSPC image of the Trifid Nebula}
\label{trixray}
\begin{center}
\begin{tabular}{lllcccl}
\hline \hline
Name & RA (2000)  & DEC (2000)  & Count Rate & Detection ($\sigma$) & 
Log (L$_{\rm x}$) \\
\hline
RXT1 & 18:02:52.8 &   $-$23:02:18.1 &  0.0014$\pm$0.0005 &  5   & 31.12 \\
RXT2 & 18:02:39.3 &   $-$22:58:34.3 &  0.0011$\pm$0.0004 & 3.5  &31.02 \\
RXT3 & 18:02:41.18 &  $-$23:03:51.8 &  0.0014$\pm$0.0006 & 3.5  &31.12 \\
RXT4 &  18:02:36.0 &  $-$23:01:36.3 &  0.0019$\pm$0.0006 &   4  &31.05  \\
RXT5 & 18:02:35.0 &   $-$23:01:29.4 &  0.0021$\pm$0.0006 &  4.5 & 31.29\\
RXT6 &  18:02:27.9 &  $-$22:59:47.9 &  0.0016$\pm$0.0006 &  4   &31.17 \\
RXT7 & 18:02:25.4 &   $-$22:59:51.4 &  0.0010$\pm$0.0005 &   3  &30.97 \\
RXT8 & 18:02:23.35  & $-$23:01:47.0 &  0.0131$\pm$0.0013 &   8  &(see Table ~\ref{trispec})  \\
RXT9 & 18:02:21.1 &   $-$23:03:21.5 &  0.0010$\pm$0.0004 & 4    &30.97 \\
RXT10 &  18:02:12.38  &$-$22:55:37.0&  0.0012$\pm$0.0004 & 4.5  &31.05 \\
\hline
RXT11 & 18:02:34.92 &  $-$22:59:55.6 & 0.0006$\pm$0.0003 & 2.5  &30.75 \\
RXT12 &18:02:31.67 &  $-$23:02:25.6 & 0.0007$\pm$0.0003  & 2.5 &30.82  \\
\hline
\end{tabular}
\end{center}
\end{table}

\font\small=cmr10 at 8pt
\font\smalit=cmti10 at 8pt
\font\smalbf=cmbx10 at 8pt

\small
\begin{table}
\caption{ \small {2MASS Young Stellar Object and T Tauri Star 
candidates  in the Trifid Nebula$^a$}}
\label{tri2mass}
\begin{center}
\vskip -0.5truecm
\begin{tabular}{clcl|clcl}
\hline \hline
&  YSO candidates \span\omit \span\omit & & TTS candidates \span\omit 
\span\omit\\
\hline
\#  &       Designation    &  \# & Designation &  \# & Designation 
&\# & Designation \\
1   & 2MASSI J1802121-230439 &  23   & 2MASSW J1802379-230211   &1 
&2MASSI J1802158-230555   & 23   &2MASSW J1802331-230135   \\
2   & 2MASSI J1802211-230437 &  24   & 2MASSW J1802211-230234   &2 
&2MASSI J1802102-230403   & 24   &2MASSW J1802197-230136   \\
3   & 2MASSI J1801592-230406 &  25   & 2MASSI J1802441-230245   &3 
&2MASSI J1802129-230348   & 25   &2MASSW J1802281-230142   \\
4   & 2MASSW J1802139-230213 &  26   & 2MASSI J1802367-230302   &4 
&2MASSW J1802197-230136   & 26   &2MASSW J1802336-230154   \\
5   & 2MASSW J1802142-230144 &  27   & 2MASSI J1802450-230332   &5 
&2MASSI J1801575-230121   & 27   &2MASSI J1802199-230305     \\
6   & 2MASSI J1802168-230110 &  28   & 2MASSI J1802500-230334   &6 
&2MASSI J1802124-225856   & 28   &2MASSI J1802395-230343     \\
7   & 2MASSI J1801561-230009 &  29   & 2MASSI J1802361-230428   &7 
&2MASSI J1802173-225614   & 29   &2MASSI J1802211-230437     \\
8   & 2MASSI J1802044-225829 &  30   & 2MASSW J1802319-230440   &8 
&2MASSI J1802108-225532   & 30   &2MASSI J1802409-230459     \\
9   & 2MASSI J1802122-225825 &  31   & 2MASSI J1802372-230503   &9 
&2MASSI J1802205-225458   & 31   &2MASSI J1802349-230505     \\
10  & 2MASSW J1802202-225807 &  32   & 2MASSI J1802369-230656   &10 
&2MASSI J1802205-225458   & 32   &2MASSI J1802263-230730      \\
11  & 2MASSI J1802401-225721 &  33   & 2MASSI J1802470-230702   &11 
&2MASSI J1802162-225530   & 33   &2MASSW J1802242-230910      \\
12  & 2MASSI J1802377-225850 &  34   & 2MASSI J1802290-230704   &12 
&2MASSI J1802464-225546   & 34   &2MASSI J1802403-230913      \\
13  & 2MASSI J1802463-225912 &  35   & 2MASSW J1802247-230729   &13 
&2MASSI J1802473-225729   & 35   &2MASSI J1802384-230934      \\
14  & 2MASSI J1802433-225931 &  36   & 2MASSI J1802344-230806   &14 
&2MASSI J1802367-225804   & 36   &2MASSI J1803003-230134      \\
15  & 2MASSI J1802491-225945 &  37   & 2MASSI J1802339-230924   &15 
&2MASSW J1802312-225911   & 37   &2MASSI J1802467-230130      \\
16  & 2MASSI J1802503-225949 &  38   & 2MASSI J1802518-230810   &16 
&2MASSW J1802305-225929   & 38   &2MASSI J1802538-230033      \\
17  & 2MASSI J1802438-225959 &  39   & 2MASSI J1802499-230740   &17 
&2MASSW J1802228-225935   & 39   &2MASSI J1802566-230026      \\
18  & 2MASSI J1802441-230020 &  40   & 2MASSI J1802541-230736   &18 
&2MASSI J1802175-225938   & 40   &2MASSI J1803023-230022      \\
19  & 2MASSI J1802266-230036 &  41   & 2MASSI J1802473-230428   &19 
&2MASSI J1802409-225941   & 41   &2MASSI J1802490-225656      \\
20  & 2MASSW J1802226-230101 &  42   & 2MASSI J1802482-230309   &20 
&2MASSI J1802344-230102   &       &                          \\
21  & 2MASSI J1802467-230130 &  43   & 2MASSI J1802524-230227   &21 
&2MASSI J1802284-230115   &       &                          \\
22  & 2MASSW J1802271-230201 &  44   & 2MASSI J1802548-225901   &22 
&2MASSW J1802408-230131   &       &                          \\
\hline                                                           
\end{tabular}
\end{center}
\vskip -0.5truecm
\tablenotetext{a}{\small {2MASSI and 2MASSW indicate the Incremental 
Release and Working Database catalogs, respectively.}}
\end{table}

\normalsize
\clearpage

\clearpage

\begin{table}
\caption{Infrared Counterparts of the X-ray Sources Detected
in PSPC Images}
\label{triinfmatch}
\begin{center}
\begin{tabular}{llrcccccl}
\hline \hline
  RXT& 2MASS$^a$ & d$^b$ & R.A. (2000)$^c$ & Dec (2000)$^c$ &
  J$^d$ & H$^d$ & K$_s$$^d$ & Comments$^e$  \\
     &Counterparts& ($''$) & (h:m:s) & ($^\circ$:$'$:$''$) &  (mag) & (mag) &
(mag) & \\
\hline
RXT1 & YSO & 10  & 18:02:52.4  & $-$23:02:27.2  & 15.17 &  13.93 &
13.18 & -  \\
RXT2 & red & 8   & 18:02:39.7  & $-$22:58:29.0  & 17.72 &  15.06 &
13.50 & J=`fill' \\
RXT3 & red & 5   & 18:02:41.4  & $-$23:03:48.2  & 17.83 &  14.24 &
12.42 & J=`fill' \\
      & red & 13  & 18:02:40.4  & $-$23:03:58.9  & 15.58 &  12.37 &
10.90 & high psf $\Delta \chi^2$\\
RXT4 & red & 12.8 & 18:02:36.8 & $-$23:01:43.9  & 15.85 &  12.78 &
11.34 & high psf $\Delta \chi^2$\\
RXT5 &   NONE \span\omit \span\omit\\
RXT6 & TTS & 13  & 18:02:30.5  & $-$22:59:28.7  & 12.71 &  11.65 &
10.91 & -\\
RXT7 & YSO & 12  & 18:02:26.6  & $-$23:00:36.0  & 15.21 &  12.90 &
11.34 & - \\
RXT8$^f$ &  Multiple &  &\\
RXT9 & red & 6   & 18:02:21.5  & $-$23:03:19.0 & 15.04 &   13.58 &
11.87 & J=`fill' \\
RXT10 &  NONE &  &\\
RXT11$^f$ & red & 4  & 18:02:34.7  & $-$22:59:52.1  & 16.52 &  12.42 &
10.58 & J=`fill' \\
RXT12 & red & 9  & 18:02:31.3  & $-$23:02:18.2  & 15.12 &  13.17 &
11.36 & J=`fill' \\
\hline
\end{tabular}
\end{center}
\tablenotetext{a}{\small Notes for the 2MASS sources: TTS = T Tauri star; YSO =
massive young stellar
    objects; red = red 2MASS star}
\tablenotetext{b}{\small Projected distance between the X-ray and 
2MASS sources.}
\tablenotetext{c}{\small Coordinates of the 2MASS counterparts.}
\tablenotetext{d}{\small J, H and K$_s$ magnitudes of the 2MASS sources.}
\tablenotetext{e}{\small J=`fil' means  the J band photometry is 
measured in band-filled within the aperture, indicating their
photometric uncertainties are large. `high psf $\Delta \chi^2$' means
the goodness of the fit of a point  spread function is high, indicating
the sources are either extended or unresolved double sources.}
\tablenotetext{f}{\small RXT8 is HD 164492 (O7  star) and/or YSO (TC1 
in CLC98),  and
RXT11 also coincides with HD 313596 (B8 star).}
\end{table}

\begin{table}
\caption{Spectral fit results from simultaneous fit using ASCA/PSPC}
\label{trispec}
\begin{center}
\begin{tabular}{lcc}
\hline \hline
   Parameters    & Trifid   \\
\hline
  ISM N$_H$ (10$^{21}$ cm$^{-2}$)  &  $\equiv$ 3\\
%\multicolumn{4}{|l|}{ASCA/ROSAT data}
%\multicolumn{2}{l}{Hot component} \\
T$_{2}$  & 3.3$\pm0.6$ keV (= 3.9$\times 10^7$ K)\\
Log EM$_{2}$ (cm$^{-3}$) & 55.8 \\
Wind N$_H$, hot component (10$^{21}$ cm$^{-2}$) & 2.7$^{+2.3}_{-2.0}$\\
T$_{1}$  & 0.14$^{+0.06}_{-0.04}$ keV (= 1.2$\times 10^6$ K) \\
Log EM$_{1}$ (cm$^{-3}$) &  57.4  \\
Wind N$_H$, 1 component (10$^{21}$ cm$^{-2}$)  &5.9$^{+1.1}_{-0.9}$\\
\hline
flux &  2$\pm$0.5 $\times$10$^{-12}$\\
unabsorbed flux  &   8$\pm$2$\times$10$^{-11}$   \\
Luminosity (erg s$^{-1}$)  & 2.5$\times$10$^{34}$ [d(kpc)/(1.67 kpc)]$^2$ \\
\hline
Fe K Line Central energy (keV) & 6.65$^{+0.1}_{-0.2}$   \\
Fe K Equivalent Width (keV)   & 1.7$\pm0.3$   \\
Fe K Flux  (photons s$^{-1}$ cm$^{-2}$) & 6($\pm$0.4)$\times$10$^{-4}$  \\
\hline \hline
\end{tabular}
\end{center}
\end{table}

%\begin{references}

%\def\pp{\parshape 2 0truecm 23truecm 0.5truecm 22.5truecm}
%\def\jref#1;#2;#3;#4;#5 {\par\noindent \pp#1 #2, #3, #4, #5. \par}
% author (date); title; journal; volume; page

\clearpage
%\noindent {\bf  Figure Captions}
%Fig. 1

\begin{figure}
%Fig.  1
\figcaption{The PSPC X-ray image and detected sources (RXT)
are marked with numbers.
The count rates for each source
are listed in Table ~\ref{trixray}.
Known sources are also marked with the labels. }
\end{figure}

%Fig. 2a
\begin{figure}
\figcaption{
%(a). The PSPC X-ray image and contours of the Trifid Nebula (J2000),
%and known sources are marked as circles with the labels.
%  The contours
%  are from 1.35 to 23.0 count (15$''$pixel$)^{-1}$. The
%  difference between adjacent contours is a multiplicative factor of 1.33.
PSPC X-ray contours superimposed on an H~$\alpha$ image of the Trifid
Nebula (courtesy of Dr. Winkler).
The optical image was obtained on 4 July, 1994 (UT) from the Burrell Schmidt
telescope of Case Western Reserve University, through a 25 \AA
~bandpass H~$\alpha$ filter.  The total exposure time
is 1800 s, and the scale is
2.0$''$pixel$^{-1}$.
The strongest X-ray peak is at the O star, HD 164492.}
\end{figure}

%Fig. 3
\begin{figure}
\caption{ (a) Near-infrared JHK$_s$ color-color diagram:
location in color-color diagram is used to
determine T Tauri-stars and massive young stellar objects.
Normal stars (dots), T Tauri-stars (filled circles),
and massive young stellar objects (diamonds) are shown with
different symbols.
The 2MASS counterparts of X-ray sources are marked with crosses;
the sources above the extinction curve have high photometric
uncertainties. They are 2MASS-red stars in Table ~\ref{triinfmatch}
(see the text for details).
Extinction vector is shown for 5 magnitude as the thick line labeled A$_v$. 
The thick curves are intrinsic colors of giant and dwarf stars. 
(b) Diagram of H-K$_s$ vs. K$_s$.
T Tauri stars and young massive protostars
fall below (younger age) the  PMS stars in this diagram.
The symbols are the same as those in (a).
(c) T Tauri stars (circles)
and massive young stellar objects (diamonds), X-ray sources 
(crosses), and known embedded young stellar objects
(triangles:from  Cernicharo et al. 1998) are marked on Optical image.
(d) 2MASS three color image (blue, green and red for J, H, and 
K$_s$, respectively).
The symbols are the same as those in (c). The diffuse, blue emission 
is probably P $\beta$ (in the J band)
from the HII region.}
\end{figure}

%Fig. 4
\begin{figure}
\caption{ASCA SIS image (gray scale ranges 0.7-4.5
counts (6.4$''$pixel)$^{-1}$)
  superposed on ROSAT contours.}
\end{figure}

%Fig. 5
\begin{figure}
\caption{(a) GIS1 and GIS2 spectra of the Trifid Nebula with its best-fit
of a two-temperature thermal model with additional warm absorbing media.
Hard emission and Fe XXV line appear in the spectra.
(b) SIS0 and SIS1 and PSPC spectra with the best-fits.
(c) Each of two temperature components
($\sim$1.2$\times$10$^6$ K and 3.9$\times$10$^7$ K) is  marked on the 
SIS0 spectrum.}
\end{figure}

  \clearpage

\psfig{figure=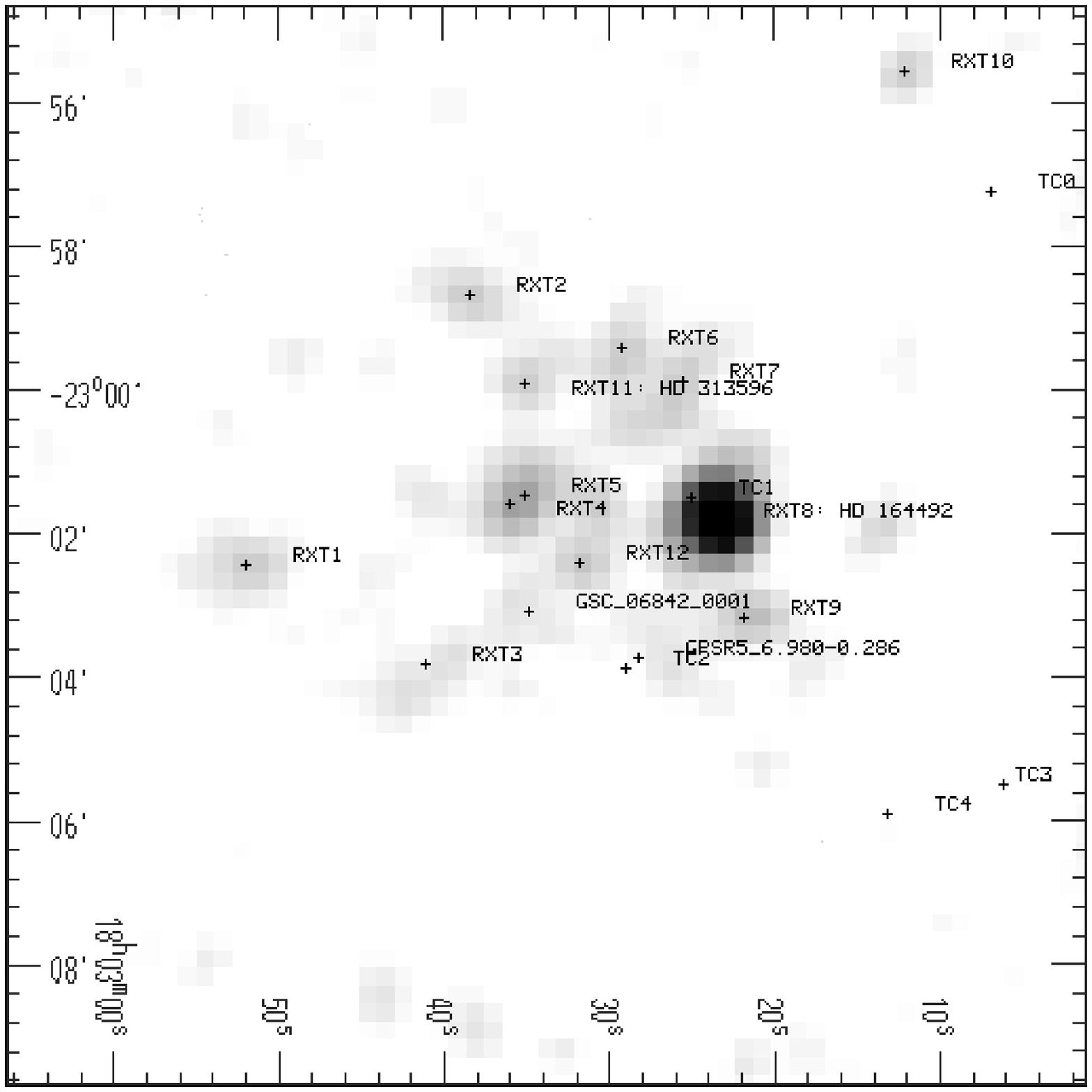,height=15.5cm}
  {Fig. 1}
  \vskip 2truecm

  \psfig{figure=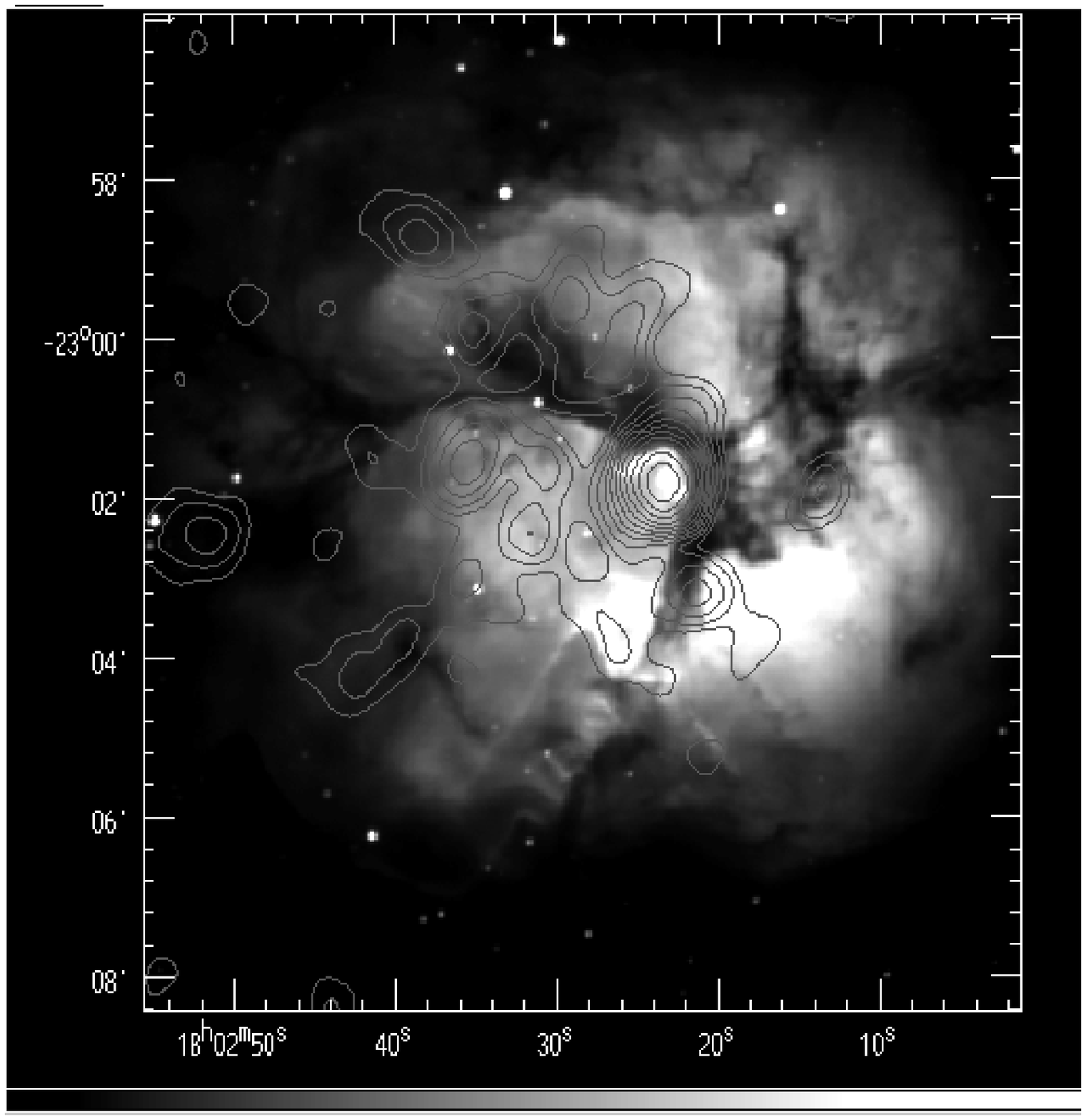,height=20.5cm}
  \vskip -3truecm
  {Fig. 2}

  \clearpage

\psfig{figure=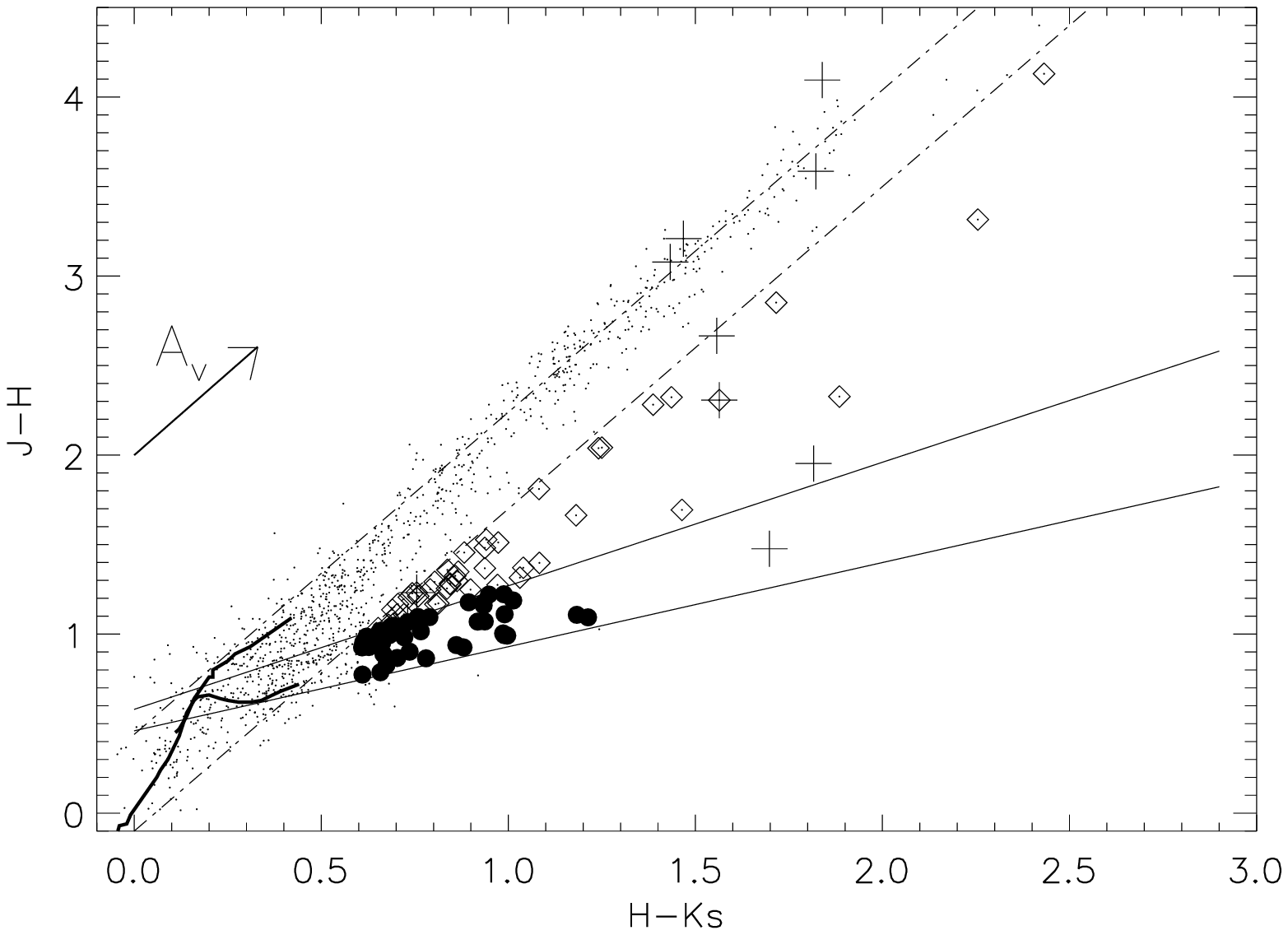,width=15.5cm}
  {Fig. 3a}
  \vskip 2truecm

  \clearpage

  \clearpage

\psfig{figure=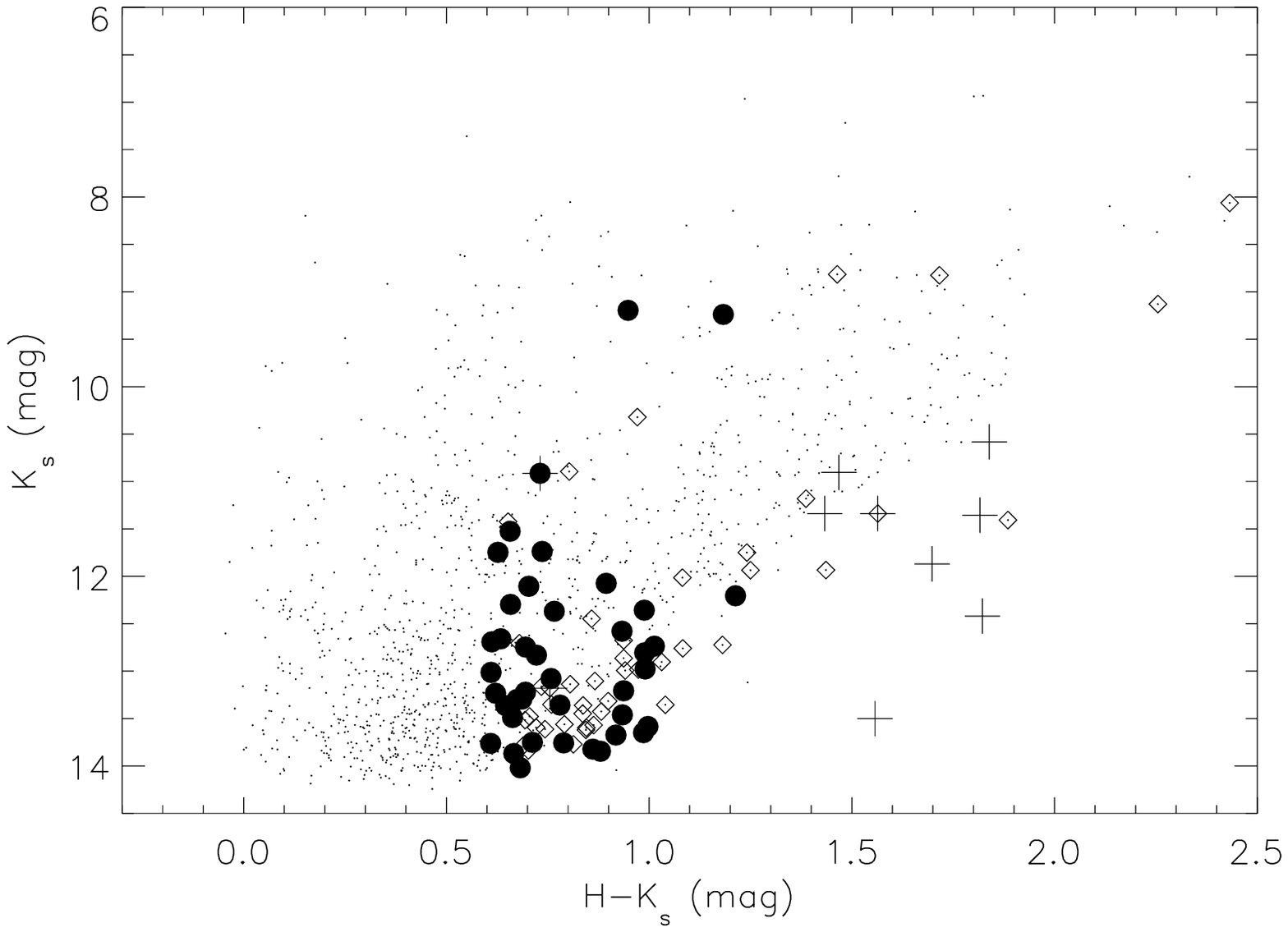,width=15.5cm}
  {Fig. 3b}

\psfig{figure=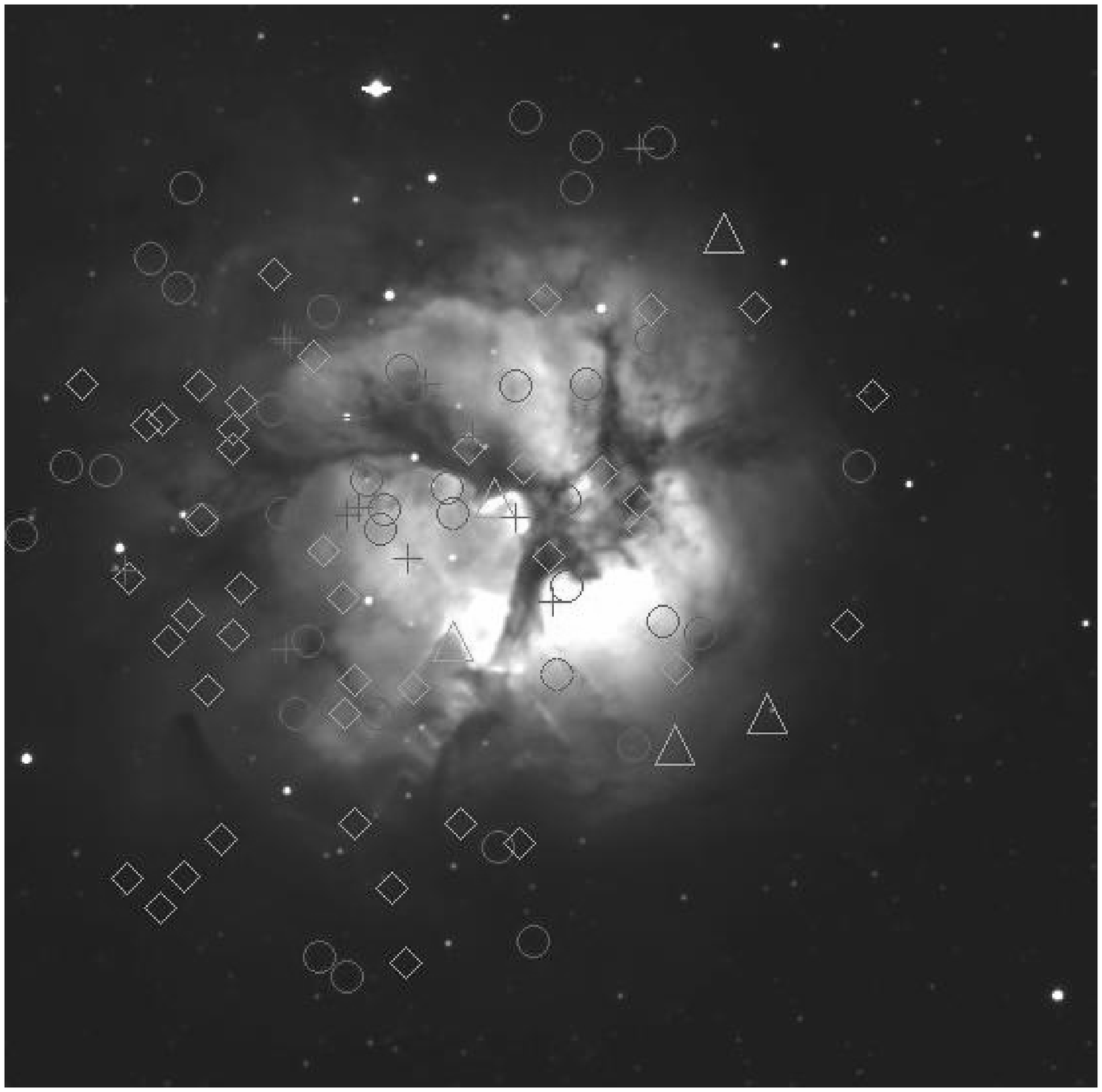,height=15.5cm}
  {Fig. 3c}
  \vskip 2truecm

\clearpage

  {Fig. 3d is a color figure (see separate page)}

\clearpage

\psfig{figure=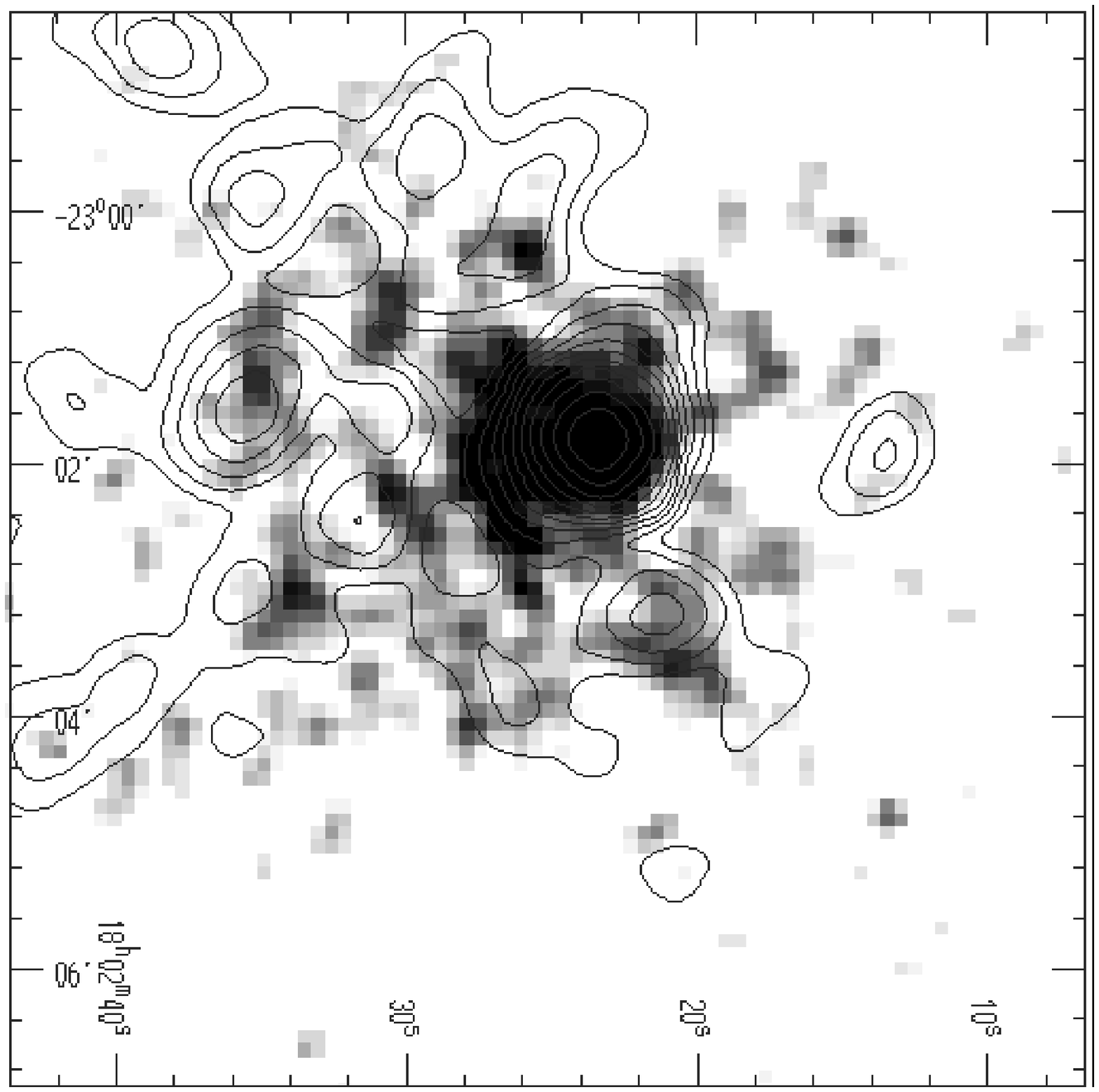,height=12.0cm}
  %\vskip 14truecm
  {Fig. 4}
  %\vskip 2truecm

\clearpage

\psfig{figure=f5a.ps,height=8cm,width=13truecm,angle=270}
  {Fig. 5a}
\vskip 0.5truecm

\psfig{figure=f5b.ps,height=8cm,width=13truecm,angle=270}

{Fig. 5b}
\vskip 0.5truecm

\psfig{figure=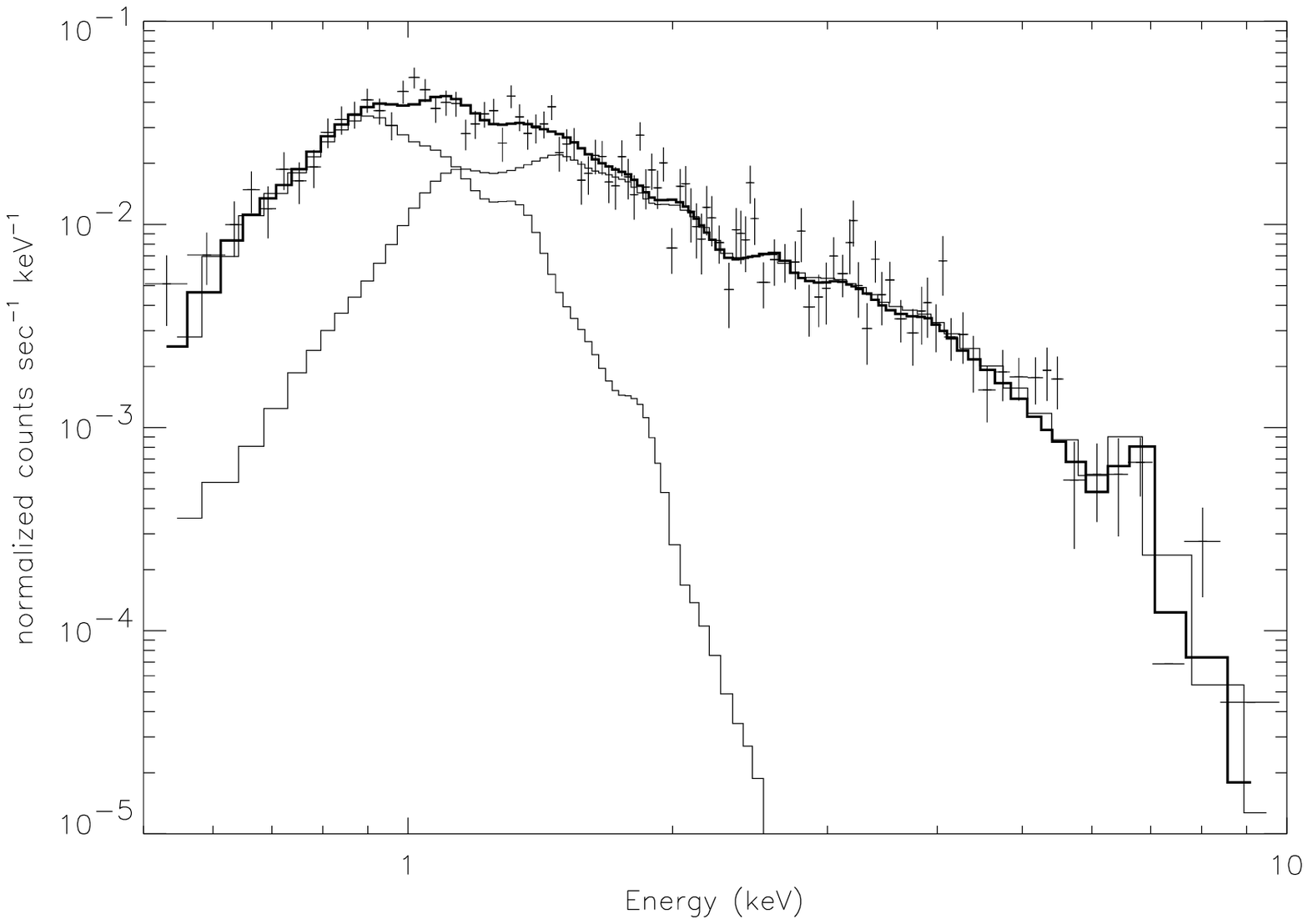,height=8cm,width=13truecm}
{Fig. 5c}

\end{document}